\documentstyle[preprint,floats,pra,aps,psfig]{revtex}

\newcommand{\doublespace}{
    \renewcommand{\baselinestretch}{1.6}\large\normalsize}

\newcommand{\bce}{\begin{center}}
\newcommand{\ece}{\end{center}}
\newcommand{\be}{\begin{equation}}
\newcommand{\ee}{\end{equation}}
\newcommand{\bea}{\vspace{0.25cm}\begin{eqnarray}}
\newcommand{\eea}{\end{eqnarray}}

\def\PLA{{Phys. Lett.}  A }
\def\PLB{{Phys. Lett.}  B }
\def\PRL{{Phys. Rev. Lett.} }
\def\PRA{{Phys. Rev.} A }

\def\PRD{{Phys. Rev.} D }

\doublespace

\begin{document}

\title{{\LARGE {\bf Can experimental tests of Bell inequalities performed
with pseudoscalar mesons be definitive? }}}

\doublespace

\author{M.Genovese \footnote{ genovese@ien.it. Tel. 39 011 3919234, fax 39
011 3919259}, C.Novero}
\address{Istituto Elettrotecnico Nazionale Galileo Ferraris, Str. delle Cacce 
91,\\I-10135 Torino, }
\author{E. Predazzi}
\address{Dip. Fisica Teorica Univ. Torino and INFN, via P. Giuria 1,
I-10125 Torino }
\maketitle

\vskip 1cm
{\bf Abstract}
\vskip 0.5cm 
We discuss if experimental tests of Bell inequalities performed with
pseudoscalar mesons (K or B) can be definitive. Our conclusion is that this
is not the case, for the efficiency loophole cannot be eliminated.

\vskip 1.5cm

\vskip 2cm
PACS: 13.20.Eb, 03.65.Bz

Keywords: neutral kaons,  Bell inequalities, Pseudoscalar mixing,
non-locality, hidden variable theories 

\vspace{8mm}

The idea that Quantum Mechanics (QM) could be an incomplete theory,
representing a statistical approximation of a complete deterministic theory
(where observable values are fixed by some hidden variable) appeared
already in 1935 thank to the celebrate Einstein-Podolsky-Rosen paper
\cite{EPR}. 
 
A fundamental progress in discussing possible extensions of QM was the
discovery of Bell \cite{Bell} that any realistic Local Hidden Variable LHV
theory must
satisfy certain inequalities which can be violated in QM leading in
principle to a possible  experimental test of the validity of QM as
compared to LHV.

Since then, many interesting experiments (in practice all based on
entangled photon pairs) have been devoted to a test of Bell inequalities
\CITE{Mandel,asp,franson,type1,type2}, leading to a substantial agreement
with standard quantum mechanics (SQM) and strongly disfavouring LHV
theories, but, so far, no experiment has yet been able to exclude
definitively such theories. In fact, so far,  one has always been forced to
introduce a further additional hypothesis \CITE{santos}, due to the low
total detection efficiency, stating that the observed sample of particle
pairs is a faithful subsample of the whole. This problem is known as { \it
detection or efficiency loophole}. The search for new experimental
configurations able to overcome the detection loophole is of course of the
greatest interest. 

In the 90's big progresses in this direction have been obtained by using
parametric down conversion (PDC) processes for generating entangled photon
pairs with high angular correlation. 
The generation of entangled states by parametric down conversion (PDC) has
replaced other techniques, such as the radiative decay of excited atomic
states, as it was in the celebrated experiment of A. Aspect et al.
\CITE{asp}, for it overcomes some former limitations. Many interesting
experiments have been realised using such a technique. The first
experiments had, by construction, a limited total efficiency
\cite{franson,type1,ou} and were far from eliminating the detection
loophole    \cite{santos}.
More recently, an experiment, based on  Type II PDC \cite{type2}, has
obtained a much higher total efficiency than the previous ones (around
$0.3$), which is, however, still far from the required value of $0.81$.
Also, some recent experiments studying equalities among correlations
functions rather than Bell inequalities \cite{dem} are far from solving
these problems \cite{garuccio}. A large interest remains therefore for new
experiments increasing total quantum efficiency in order to reduce and
finally overcome the efficiency loophole. 

Some years ago, a very important theoretical step in this direction was
performed recognising that, while for maximally entangled pairs a total
efficiency larger than to 0.81 is required to obtain an efficiency-loophole
free experiment, for non maximally entangled pairs this limit is reduced to
0.67 \cite{eb} (in the case of no background). An experiment addressed to
test Bell inequalities using non-maximally entangled photon pairs has been
recently realised \cite{nos}. Work is in progress for obtaining an
efficiency above $0.67$ with this kind of set-ups.

Even if relevant progresses toward the elimination of the detection
loophole have been obtained using entangled photon pairs, nevertheless the
total efficiency is strongly dominated by the quantum efficiency of
photodetectors.
Nowadays efficiencies for commercial photodetectors are around 70 \%.
Prototypes already reach much higher efficiencies \cite{Yam}, but at the
prize of  high background which also limits the possibility of a loophole
free test \cite{Tho}. Thus, in summary, the use of entangled photon pairs
has led to very important tests of Bell inequalities, but at the moment
does not allow to eliminate the detection loophole.

On the other hand, a recent experiment \cite{Win} performed using Be ions
has reached very high efficiencies (around 98 \%), but in this case the two
subsystems (the two ions) are not really separated systems and the test
cannot be considered a real implementation of a detection loophole free
test of Bell inequalities \cite{Vai}, even if constitutes a relevant
progress in this sense.  

Even if little doubts remain on the validity of the standard quantum
mechanics, considering the fundamental importance of the question, the
search for other experimental schemes for a definitive test of Bell
inequalities is therefore of the largest interest.

In the last years many papers have been devoted to study the possibility of
realising such a test by the use of pseudoscalar meson pairs as $K \bar{K}$
or $B \bar{B}$. If the pair is produced by the decay of a particle at rest
in the laboratory frame (as the $\phi$ at Daphne), the two particles can be
easily separated to a relatively large distance allowing an easy space-like
separation of the two subsystems and permitting an easy elimination of the
space-like loophole, i.e. realising two completely space-like separated
measurements on the two subsystems (where the space-like separation must
include the setting of the experimental apparata too). A very low noise is
expected as well.

The idea is to use entangled states of the form:
\bea
|\Psi \rangle = { | K_0 \rangle | \bar K_0 \rangle  - | \bar K_0 \rangle |
K_0 \rangle \over \sqrt{2} } = & \cr
= { | K_L \rangle |  K_S \rangle  - |  K_S \rangle | K_L \rangle \over
\sqrt{2} } & \cr
\label{psi}
\eea

Claims that these experiments could allow the elimination of the detection
loophole for the high efficiency of particles detectors, have also been
made. In this letter we study critically this statement.

The main caveat derives from the fact that in any experimental test
proposed up to now one must tag the $P$ or $\bar{P}$ trough its decay. This
requires the selection of $\Delta S = \Delta Q$ semileptonic decays, which
represent only a fraction of the total possible decays of the meson, e.g
$BR(K^0_S \rightarrow \pi^{+} e^{-} \nu_e) = (3.6 \pm 0.7)
10^{-4}$,$BR(K^0_L \rightarrow \pi^{+} e^{-} \nu_e) = 0.1939 \pm 0.0014$,
$BR(K^0_L \rightarrow \pi^{+} \mu^{-} \nu_{\mu}) = 0.1359 \pm 0.0013 $,
$BR(B^0 \rightarrow  l^{+} \nu_{l} X) = 0.105 \pm 0.008 $ \cite{PDB}).
 Furthermore, experimental cuts on the energies of the decay products  will
inevitably reduce further this fraction and part of the pairs could be lost
by decays occurring before the region of observation. Finally, most of
these proposals involve the regeneration phenomenon, which introduces
further strong losses. Thus, one is led to subselect a fraction of the
total events. As one cannot exclude a priori  hidden variables related to
the decay properties of the meson, one cannot exclude the sample to be
biased and thus the detection loophole pops out again.
This is in analogy to the photon experiments, where the detection loophole
derives by the fact that one cannot exclude  losses  related to the values
of hidden variables which determine if the photon passes or not a
polarisation (or another) selection.  Namely, in a local realistic model
its properties are completely specified by the hidden variables. Also
decays, in a deterministic model, can happen according to the values of the
hidden variables (both in a deterministic or in a probabilistic way). Thus,
states with different hidden variables can decay in different channels,
with the condition that the branching ratios {\it averaged } on the hidden
variables distribution reproduce the quantum mechanics predictions. 

For the experiments based on Bell inequalities measurements \cite{BellK},
the limits discussed before for the total efficiency remain valid. As the
total branching ratio in  $\Delta S = \Delta Q$ semileptonic decays is much
smaller than 0.81 (the eventual use of non-maximally entangled states,
lowering the efficiency threshold to 0.67, does not change the situation),
this inevitably implies that a loophole free test of Bell inequalities
cannot be  performed in this case (even neglecting other problems
\cite{gh}). It must be noticed that this problem does not appear in Ref.
\cite{BF}, however other additional hypotheses are needed (see Eq. 15 and
discussion after Eq. 18 of \cite{BF}), and thus this proposal does not
allow a general test of HVT as well.

It must also be noticed  that the only observation of interference between
the two term of the entangled wave function, Eq. \ref{psi}, as in Ref
\cite{CLEO}, does not exclude  general HVT, for this feature can be
reproduced in an general class of local realistic theories. 
 
Let us then consider other proposals  not based on a  Bell inequalities
measurement. Two proposals of this kind have been recently advanced by F.
Selleri and others concerning a  $K \bar{K}$ \cite{Sel1} or  a $B \bar{B}$
\cite{Sel2} system respectively.

Let us begin analysing the $K \bar{K}$ case (the $B \bar{B}$ one follows
with small modifications.)

In the model of Ref. \cite{Sel1}, the $K \bar{K}$ pair is
local-realistically described by means of two hidden variables, one
($\lambda_1$) determining a well defined CP value, the other ($\lambda_2$)
a well defined strangeness $S$ value for the $K$ (and related to this for
the $\bar K$). This second variable cannot be a time independent property,
but is subject to sudden jumps. If locality must be preserved the time of
this jump must already be fixed ab initio by a hidden variable (which
represents the real second hidden variable of the model) and the two
subsystems must not influence each other while they are flying apart,
namely $\lambda_2$ is not the true hidden variable, but a parameter driven
by the true hidden variable (see appendix of Ref. \cite{Sel1}).

Let us denote by $K_1$ the state with CP=1, S=1, $K_2$ the state with CP=1,
S=-1,
$K_3$ the state with CP=-1, S=1 and
$K_4$ the state with CP=-1, S=-1.

The initial state can be, with probability $1/4$, in anyone of the states
$CP= \pm 1$, $S=\pm 1$. Each of these pairs give, in the local-realistic
model (LRM), a certain probability of observing a $\bar K_0 \bar K_0$ pair
at proper times $t_a$ and $t_b$ ($\ne t_a$) of the two particles. These
probabilities are (in a somehow simplified form, see eq. 62-70 of Ref.
\cite{Sel1}):
\bea
P_1[t_a,t_b]=[E_S(t_a) Q_-(t_a) - \rho(t_a)] \cdot E_L(t_a) p_{43}(t_b |
t_a) & \cr
P_2[t_a,t_b]=[E_S(t_a) Q_+(t_a) + \rho(t_a)] \cdot E_L(t_a) p_{43}(t_b |
t_a) & \cr
P_3[t_a,t_b]=[E_L(t_a) Q_-(t_a) + \rho(t_a)] \cdot E_S(t_a) p_{21}(t_b |
t_a) & \cr
P_4[t_a,t_b]=[E_L(t_a) Q_+(t_a) - \rho(t_a)] \cdot E_S(t_a) p_{21}(t_b |
t_a) \, \,
\label{P} 
\eea
corresponding to an initial state with $K_1$ on the left and $K_4$ on the
right,
$K_2$ on the left and $K_3$ on the right, $K_3$ on the left and $K_2$ on
the right and $K_4$ on the left and $K_1$ on the right respectively.

In Eq. \ref{P}, we have introduced $ E_S(t) = exp(- \gamma_S t)$ and $
E_L(t) = exp(- \gamma_L t)$, where $\gamma_{S}=(1.1192\pm 0.0010) 10^{10}
s^{-1}$
and $\gamma_{L}=(1.934 \pm 0.015) 10^{7} s^{-1}$ denote the decay rate of
$K_S$ and $K_L$ \cite{PDB}. 
$Q_{\pm} ={ 1 \over 2}  \left[ 1 \pm {2 \sqrt{E_L E_S} \over E_L + E_S}
\cos( \Delta m t) \right]$, where $\Delta m = M_L -M_S = (0.5300 \pm
0.0012) 10^{10} s^{-1}$.
Furthermore, we have defined $ p_{21}(t_b | t_a) = E_S^{-1}(t_a)
[p_{21}(t_b | 0) - p_{21}( t_a | 0)  \cdot E_S(t_b-t_a)]$
and
$ p_{43}(t_b | t_a) = E_L^{-1}(t_a) [p_{43}(t_b | 0) - p_{43}( t_a | 0)
\cdot E_L(t_b-t_a)]$
where $ p_{21}(t | 0) = E_S(t) Q_-(t) - \rho(t)$ and $ p_{43}(t | 0) =
E_L(t) Q_-(t) + \rho(t)$.
Finally, $\rho(t)$ is a function not perfectly determined in the model (see
discussion in Ref. \cite{Sel1}), but which is limited  by
\bea
-E_S Q_+ \le \rho \le E_S Q_- \cr
-E_L Q_- \le \rho \le E_L Q_+ \cr
\label{ro}
\eea

If the total efficiency is 1,  the LRM probability of observing a $\bar K_0
\bar K_0$ pair is given by the sum of the four probabilities of Eq. \ref{P}
multiplied for $1/4$. It is rather different from the quantum mechanical
prediction and thus represents a good test of the LRM (see fig. 1, where  $
P[\bar{K_0}(t_a), \bar{K_0}(2 t_a)]$ is reported in analogy to Table 1 of
Ref. \cite{Sel1}). Nevertheless, when the total efficiency is lower than 1,
the different probabilities can contribute in different way as the hidden
variables, which determines the passing or not the test, could  also be
related to the decay properties of the meson pair.  As discussed
previously, the specific property of the meson is not being or not a $\bar
K_0$ at a certain proper time, but the hidden variables values characterise
it completely, and thus, in principle, even its decay properties. 
If this is the case, different coefficients $a_i$   can multiply the four
probabilities. One has therefore:
\bea
P[\bar{K_0}(t_a), \bar{K_0}(t_b)]= 1 / 4 \cdot \left[ a_1 P_1 [t_a,t_b] +
a_2 P_2 [t_a,t_b] + a_3 P_3 [t_a,t_b] + a_4 P_4 [t_a,t_b] \right]
\eea

The freedom of the choice of this parameters allow to reproduce the quantum
mechanical prediction.
In figure 2 we report the case corresponding to a total efficiency of 0.3
(other values can be obtained by scaling): the lowest limit curve of local
realism can easily reproduce or be lower than the quantum mechanics
prediction when different weights multiply the four probabilities, due to
different branching ratios for the 4 cases.
As an example, the curve corresponding to LRM with weights $a_1=1$,$a_2=
0.07$,$a_3= 0.03$ and $a_4=0.1$ is shown. For the sake of simplicity, in
this example the values of $a_i$ are chosen such that their sum is the
total efficiency, but the value of the $a_i$ are substantially very little
constrained as they can depend on the time: only the total fraction of
observed decays should be reproduced.
Furthermore, the result is obtained with  $\rho=0$. Thanks to the large
arbitraryness of $\rho$, other choices would  even allow to reproduce
easier  the quantum mechanics result (as we have calculated for a lot of
different choices of $\rho$). It must be emphasised that the LRM curve in
fig.2  is not a fit to the quantum mechanics curve, but only a simple
choice showing the effect we are discussing. Considering the large
arbitrary of a general HVT, our purpose is only to show a counterexample,
which proves that observation of the curve predicted by SQM cannot exclude
{\it every } HVT. Of course, as all Bell inequalities tests performed up to
now, a result in agreement with SQM will further reduce the space for the
existence of a realistic HVT, even if unable to obtain the general result
of eliminating the possible existence of HVT.
  
 Let us notice that situation 1 and 4, 2 and 3 are symmetric under the
exchange of left and right, however the decay probabilities need not to be
the same. Furthermore, let us emphasise again that, in principle, inside
this model the coefficients could even be function of the time $a_i(t)$,
like the value for the hidden variable $\lambda_2$. This property of course
makes  easier  to reproduce the quantum mechanics prediction. 
Thus, when only a subsample is selected  (semileptonic  $\Delta S = \Delta
Q$ decays must be observed for tagging a $\bar K_0 \bar K_0$ state and cuts
must be introduced) the result of this analysis shows that detection
loophole appears also in this case. Of course we are not discussing how the
local realistic  model should be, or if its complexity makes it
unpalatable, we are simply investigating if every local realistic model
could be excluded without any doubt by such experiments and we conclude
that one is focused  to introduce the additional hypothesis that the
observed sample is unbiased concerning the hidden variables values.
Even if for some specific function $\rho(t)$, it could not be possible, for
time independent decay properties, to obtain 
a perfect agreement between LRM and SQM, for the  function $\rho(t)$ is not
fixed for a general LRM, this does not modify our conclusions.

 Exactly the same considerations apply for what concerns  the $B_0
\bar{B_0}$ case.
A first set of papers \cite{BF} consider the possibility to test SQM
measuring the term deriving by interference between the two terms in the
entangled wave function of Eq. \ref{psi}. However, this effect is also
reproduced in any reasonable HVT \cite{Sel2} and thus cannot be considered
a general test of local realism, but can only allow to eliminate some
specific class of hidden variable theories.
As the model of Ref. \cite{Sel2} is, with small changes due to the same
decay time for both the CP eigenstates, equivalent to the one we have
discussed in the previous paragraph, the same conclusions hold. 

In conclusion, even if tests of local realism using pseudoscalar mesons
represent an interesting new way of investigating quantum non-locality in a
new sector, it still appears to us that none of the proposed schemes
permits a conclusive test of local realism, for the impossibility of
eliminating the detection loophole.

\vfill \eject

\vskip 1cm
{\bf Figures Captions}

Fig.1   The SQM and the minimal  LRM predictions for $ P[\bar{K_0}(t_a),
\bar{K_0}(2 t_a)]$ ($\rho=0$). The minimal LRM  is largely above the SQM
prediction. For the sake of completeness we report  the four probabilities
$P_i$ in function of the proper time $t_a$ ($t_b = 2 t_a$) as well. The
dashing of the curves diminishes in this order.

Fig.2 The SQM (dashed) and the minimal LRM, $\rho=0$, (thick) predictions
for $ P[\bar{K_0}(t_a), \bar{K_0}(2 t_a)]$ keeping into account a total
detection efficiency of 0.3. With the choice $a_1=1, a_2=0.07, a_3=0.03,
a_4= 0.1$ the two curves substantially coincides. For other choices of the
parameters the lower bound of LRM can be easily taken largely under SQM
prediction.
\end{document}